# Phonon-assisted electronic states modulation of few-layer PdSe$_2$ at terahertz frequencies

Ziqi Li[1,†], Bo Peng[2,†], Miao-Ling Lin[3], Yu-Chen Leng[3], Bin Zhang[1], Chi Pang[1], Ping-Heng Tan[3,*], Bartomeu Monserrat[2,4,*], and Feng Chen[1,*]

[1]School of Physics, State Key Laboratory of Crystal Materials, Shandong University, Shandong, Jinan, 250100, China.

[2]Cavendish Laboratory, University of Cambridge, J. J. Thomson Avenue, Cambridge CB3 0HE, United Kingdom.

[3]State Key Laboratory of Superlattices and Microstructures, Institute of Semiconductors, Chinese Academy of Sciences, Beijing 100083, China.

[4]Department of Materials Science and Metallurgy, University of Cambridge, 27 Charles Babbage Road, Cambridge CB3 0FS, United Kingdom.

†These authors contributed equally to this work.

*E-mails: drfchen@sdu.edu.cn; bm418@cam.ac.uk; phtan@semi.ac.cn.

**ABSTRACT**

Information technology demands high-speed optoelectronic devices, but going beyond the one terahertz (THz) barrier is challenging due to the difficulties associated with generating, detecting, and processing high-frequency signals. Here, we show that femtosecond-laser-driven phonons can be utilized to coherently manipulate the excitonic properties of semiconductors at THz frequencies. The precise control of the pump and subsequent time-delayed broadband probe pulses enables the simultaneous generation and detection processes of both periodic lattice vibrations and their couplings with electronic states. Combining ultralow frequency Raman spectroscopy with first-principles calculations, we identify the unique phonon mode-selective and probe-energy dependent features of electron-phonon interactions in layered $PdSe_2$. Two distinctive types of coherent phonon excitations could couple preferentially to different types of electronic excitations: the intralayer (4.3 THz) mode to carriers and the interlayer (0.35 THz) mode to excitons. This work provides new insights to understand the excited-state phonon interactions of 2D materials, and to achieve future applications of optoelectronic devices operating at THz frequencies.

**INTRODUCTION**

Fascinating quantum phenomena in atomically thin 2D materials have attracted tremendous attention from fundamental research to industry owing to their various applications in multiple areas[1]. One of the most interesting features is the reduced screening that leads to enhanced Coulomb interactions compared with their 3D counterparts[2]. For example, tightly bound electron-hole pairs, known as excitons, can form in 2D semiconductors with binding energies of hundreds of millielectronvolts (about an order of magnitude larger than those of typical 3D semiconductors)[3]. Phonons are another quasiparticle in condensed matter systems, and their interaction with other phonons, electrons, or photons is of fundamental significance to understand physical properties ranging from mobility to superconductivity. The electron–phonon interaction is typically studied at the electronic ground state using Raman spectroscopy[4], whereas excited-state phonon dynamics remains less understood. It is therefore critical to develop a deeper understanding on how different phonon modes interact with other quasiparticle excitations such as excitons.

Optical driving has emerged as a powerful method to generate novel out-of-equilibrium transient phases of quantum matter that are not available at equilibrium. A particularly useful tool to gain knowledge of nature's speediest processes are ultrafast pump–probe experiments, in which the non-equilibrium states are triggered by light pulses and the temporal evolution is subsequently monitored by appropriately-delayed probe pulses. A variety of light-driven excited states have been explored in 2D layered materials, including carrier relaxation dynamics[5-7], superconductivity[8], the anomalous Hall effect[9], charge density waves[10], chiral phonons[11], and spin-valley dynamics[12-14]. In the phonon sector, the impulsive force provided by light pulses can initiate coherent phonon oscillations, in which individual phonon modes are selectively excited and lead to atomic motions that keep pace with their neighbors. For example, vibrational wavepackets for the radial breathing mode and the G mode have been observed in carbon nanotubes[15] and 47.4 THz coherent phonons of in-plane carbon stretching mode have been reported in graphite[16]. In layered quasi-2D materials, these periodic vibrations can induce relative atomic displacements and strains that are associated to either interlayer (shear or breathing)[17-18] or intralayer[19,20] modes. In the electronic sector, the photoexcited free carriers can modify the screening properties of the material, providing an opportunity to shift the conduction band (CB) and valence band (VB) relative to each other[21-25]. This provides an avenue for manipulating the bandgap of the material in an all-optical way without any chemical modification.

Bringing the electronic and phonon sectors together also provides opportunities for applications. Future information technologies call for new devices that can operate at frequencies above 1 THz, or one trillion cycles per second. These high frequencies are particularly useful: they can carry more information (both classical and quantum) than lower frequency signals and can open up new options for applications such as actuators, radars, ultrafast data recording, processing, and communications systems. However, the high-frequency performance of switches or modulators is still a challenge, with a theoretical limitation of the modulation speed to the gigahertz (GHz) regime[26,27]. With the assistance of the strain induced by coherent phonons, one can expect to go beyond the traditional limit and manipulate the transient physical properties (*e.g.*, optical, electronic, and topological properties) that are sensitive to lattice symmetry at frequencies comparable to the terahertz atomic motions. For instance, ultrashort light pulses have been employed to achieve an ultrafast topological switch based on the shear phonon oscillations of few-layer $WTe_2$ at a frequency of 0.24 THz[28]. Despite these promising developments, the direct demonstration of terahertz-frequency modulation of the electronic bandgap is still missing in current material systems with limited probe energy range.

To address this challenge, we report on an ultrafast modulator of the electronic bandgap of few-layer palladium diselenide operating at terahertz frequencies, taking advantage of electronic (bandgap renormalization) and phonon (field-driven oscillations) effects simultaneously. Combining broadband pump-probe experiments and first-principles calculations, we reveal the dominant roles of screening in the bandgap renormalization as well as interactions with two types of coherent phonons that preferentially couple to electrons and excitons and operate at 4.3 THz and 0.35 THz, respectively. Overall, the simultaneous interplay of carriers, excitons, and phonons provides a comprehensive view of the non-equilibrium quasiparticle dynamics and many-body interactions in atomically thin 2D materials.

**RESULTS AND DISCUSSIONS**

**Growth and characterizations of large-area PdSe$_2$**

As a transition-metal dichalcogenide (TMD), palladium diselenide (PdSe$_2$) has emerged as a new type of layered material with distinctive physical properties such as remarkable stability in air, high carrier mobility, a negative Poisson's ratio, and pressure-induced superconductivity[29-31]. In addition, PdSe$_2$ is expected to possess abnormally strong interlayer coupling beyond the typical van-der-Waals forces found in layered materials, because its $d^2sp^3$ hybridization between the $p_z$ band of Se and the $d$ band of Pd is stronger than in other TMDs with $d^4sp$ hybridization[32]. Therefore, PdSe$_2$ offers an excellent platform to explore light-driven non-equilibrium many-body interactions that can be harnessed for application in functional devices. In this work, a wafer-scale PdSe$_2$ sample is first synthesized on a quartz substrate using the chemical vapor deposition (CVD) technique with high-purity $PdCl_2$ and Se under vacuum conditions (from SixCarbon Technology, Shenzhen). The top and side view of the crystal structure of $PdSe_2$ is shown in Fig. 1a, exhibiting a puckered pentagonal pattern with an orthorhombic lattice. The high quality of the sample is confirmed by detailed characterizations. The nanoscale surface topographic image is shown in Fig. 1b and the height profile of the sample is shown in Fig. 1c with a measured average thickness of ~2.9 nm. The height corresponds to eight layers with an interlayer distance of ~0.39 nm. According to the AFM images, the height is very uniform over the whole sample, implying uniformity of thickness and further confirming the sample quality. The chemical and stoichiometric characteristics are probed by X-ray photoelectron spectroscopy (XPS). Figure 1d and 1e show the high-resolution XPS spectrum of Se $3d$ and Pd $3d$ regions, respectively. The two characteristic peaks at 54.75 and 55.65 eV in Figure 1d are attributed to the $2d_{5/2}$ and $2d_{3/2}$ doublets of $Se^{2-}$ and the peaks located at 342 eV and 336.75 eV in Fig. 1e are originated from the $3d_{3/2}$ and $3d_{5/2}$ orbitals of $Pd^{4+}$. The atomic ratio of Pd and Se atoms can be quantitatively estimated to be near stoichiometric. Figure 1f shows the calculated electronic structure of $PdSe_2$ with an indirect bandgap of 1.1 eV using many-body perturbation theory. Figure 1g presents the ground-state absorption spectra of the $PdSe_2$ sample measured by UV-Vis-NIR Spectrophotometer (Agilent Cary 5000). The indirect bandgap is estimated to be 1.1 eV through a linear extrapolation Tauc plot (Supplementary Note 1), in good agreement with the calculated value. The spectrum features a pronounced A-exciton transition resonance near 2.23 eV, corresponding to the strong optical transition between the parallel bands in Fig. 1f. The calculated exciton peak at 2.31 eV is in reasonable agreement with the measurement.

**Ultrafast interplay between electronic and lattice excitations**

The temporal and spectral evolution of different electronic transitions in PdSe$_2$ is recorded in real time at room temperature by measuring the differential transmission contrast signal $\Delta T/T$ using ultrafast transient absorption spectroscopy (See Experimental Section for details). The sample is excited by 35-fs pulses centered at a photon energy of 3.4 eV and with a pump fluence set to 160 µJ/cm$^{-2}$. Figure 2a shows the 2D mapping of the $\Delta T/T$ signal as the function of pump-probe delay and probe photon energy. Figure 2b shows the $\Delta T/T$ spectra near A exciton resonance for the first 10 ps. Interestingly, the variation of the optical response appears immediately after photoexcitation and subsequently the sample exhibits a distinct response at different time delays. A pronounced photobleaching (PB) signal is observed when probing near the A excitonic transition resonance, which can be attributed to population-induced phase-space filling effects arising from Pauli blocking. The absorption can be obtained from the transmission spectrum with the thin-film approximation and the exciton resonance energy is extracted from the absorption spectrum through Lorentz fitting (Supplementary Note 2). Figure 2c shows the exciton energy shift ($\Delta E$) as a function of pump-probe time delay. With increasing time, the exciton resonance energy experiences a dramatic energy shift with a blueshift and redshift crossover as well as pronounced oscillatory components. The large magnitude of this laser-induced energy shift, compared to those in conventional semiconducting quantum well systems or 2D materials[22,33-37], is a signature of strong many-body interactions in few-layer PdSe$_2$. Initially, the exciton energy shifts to higher energies, which may be due to the Pauli blocking of the exciton band edge via the Burstein−Moss effect with high carrier density and the dynamic screening of excitons induced by the free carriers. Afterwards, a pronounced energy redshift occurs at about 1.5 ps. The microscopic mechanism behind the energy shift results from the competition between the quasiparticle bandgap ($E_g$) shrinkage and the exciton binding energy ($E_b$) reduction in the presence of strong carrier screening.

Optical excitation above the A exciton resonance generates an abundance of free-charge carriers at the very beginning that precedes A exciton formation. At such high excitation densities, photoexcited carriers strongly screen Coulomb interactions. The carrier-induced screening leads to a bandgap renormalization by reducing both $E_g$ and $E_b$[23]: the screening of the repulsive Coulomb interaction between electrons reduces $E_g$, whereas the screening of the attractive Coulomb interaction between electron-hole pairs reduces $E_b$. Such competition leads to the transition from blueshift to redshift. Our results indicate that the presence of strong screening due to excess photoexcited carriers plays a crucial role in the renormalization of the quasiparticle bandgap. As shown in Fig. 2d, under strong screening conditions, $E_g$ decreases much faster than $E_b$, and as a consequence the exciton resonance energy $E_{opt} = E_g - E_b$ redshifts, which is consistent with the energy redshift at later times. Such screening effects can be partially captured by hybrid functionals based on a screened Coulomb potential [38] (Supplementary Note 3).

Along with the pump-induced exciton resonance shift, we observe an ultrafast periodic modulation of optical properties in PdSe$_2$ on two different timescales. A plot of the temporal evolution of the transmission contrast $\Delta T/T$ over a broad range of probe photon energies exhibits the two regimes. The first regime is over femtosecond timescales, as demonstrated in Fig. 3a and Fig. 3b, showing that the optical response has a strong modulation of the differential transmission $\Delta T/T$ signal initiated by the ultrashort laser. As shown in Fig. 3c, the oscillatory part can be fitted to a damped sinusoidal model [$\Delta T/T(t, \lambda) = Be^{-\gamma t}\cos(\omega t + \phi)$, in which $B$ is the initial amplitude, $\gamma$ is the decay constant, $\phi$ is the phase angle (at $t = 0$), and $\omega$ is the angular frequency] with a repetition rate of 4.3 THz and a lifetime of 1.4 ps when probed at 1.71 eV. In addition to $\Delta T/T$, the oscillatory component of the time-dependent ultrafast energy shift $\Delta E$ is extracted and fitted with the damped sinusoidal model, as shown Fig. 3d. The observed modulation frequency of $\Delta E$ is also about 4.3 THz, corresponding to 4.3 trillion cycles per second.

Analogous to the time-domain visualization of the coherent vibrations on the femtosecond timescale, a second regime emerges on longer timescales, also exhibiting an exceptionally strong periodic modulation of the differential contrast signal $\Delta T/T$ (Fig. 4a and Fig. 4b). The modulation feature weakens as the probe photon energies are tuned from 2.14 to 1.71 eV. At 1.20 eV, the increased modulation intensity arises from the intraband absorption correlated to both free carriers and excitons. The oscillatory components of the time-resolved $\Delta T/T$ can be extracted and fitted after subtracting the bi-exponentially fitted population dynamics, as demonstrated in Fig 4c. The vibrational coherence profiles are accurately fitted using an exponentially damped sinusoidal model and the period is found to be 2.86 ps, corresponding to coherent phonon oscillations at a frequency of 0.35 THz, much slower than the oscillatory frequency of 4.3 THz. All the kinetic curves of the full-time range can be accurately fitted to a bi-exponential relaxation model and the best-fit $\tau_1$ and $\tau_2$ as a function of probe photon energy is demonstrated in Fig. 4d. The fast component $\tau_1$ can be due to extrinsic non-radiative energy channels such as Auger-type processes and the cooling of hot carriers to the lattice, and it is slightly dependent on the probe energy as different probe frequency would interrogates the carrier dynamics when fulfills the resonance condition. The slow relaxation time $\tau_2$ can be attributed to the exciton lifetime and phonon scattering[39]. As shown in Supplementary Fig. 4, the frequency of the oscillations can be further confirmed from the Fourier transform (FT) of the oscillatory component probed at 1.71 eV, showing two strong peaks with modulation frequency at 4.3 and 0.35 THz.

**Phonon dynamics revealed by low-frequency Raman spectroscopy**

Having established the importance of two distinct coherent oscillations at THz frequencies, we next proceed to examine their microscopic features. Recent advances in realizing ultra-narrow optical filters have enabled the measurement of ultralow frequency (typically < 60 cm$^{-1}$) Raman modes, offering a highly sensitive probe of interlayer vibrations and couplings in two-dimensional layered materials[4,40,41]. The Raman spectra are obtained in both low-frequency and high-frequency regimes simultaneously with an excitation energy of 2.33 eV, close to the A exciton, as shown in Fig. 5a. The Stokes/anti-Stokes Raman spectra under parallel (HH) and cross (HV) polarization configurations are also shown in Fig. 5b. In the high-frequency regime, due to the $C_{2v}$ symmetry of eight-layer PdSe$_2$, only the $A_1$ and $A_2$ modes are observed under the backscattering configuration, and they can be attributed to the intralayer vibrations of PdSe$_2$. Thus, the observed high-frequency modes, $A_1^{(i)}$ ($i = 0,1,2,...,5$) peaks depicted in Fig. 5a, can be assigned to the $A_1$ modes based on their polarized behaviors. The $A_1^i$ ($i = 1,2,...,5$) peaks correspond to the $P$i (i = 1,2,...,5) peaks highlighted of exfoliated PdSe$_2$, which can be attributed to the intralayer vibrations of PdSe$_2$. The intralayer vibration at 144.6 cm$^{-1}$ (4.3 THz) originating from the intralayer Se–Se bonds shows the strongest Raman intensity, implying its strong coupling to the electronic states. Besides, the interlayer layer-breathing (LB) and shear (S) modes in the ultralow-frequency regime can also be distinguished, which correspond to vibrations perpendicular and parallel to the basal plane, respectively. According to the density functional theory calculations and modified linear chain model[32], the interlayer modes at 12.5 cm$^{-1}$, 35.4 cm$^{-1}$, 48.1 cm$^{-1}$ and 51.6 cm$^{-1}$ can be assigned to LB$_{8,7}$, LB$_{8,5}$, LB$_{8,3}$, and LB$_{8,1}$, respectively, as shown in Fig. 5a, further confirming the layer number of 8 and also the high quality of CVD-grown wafer-scale PdSe$_2$. Distinct from many other 2D materials, the intensity of the interlayer LB modes in PdSe$_2$ is unusually high and is comparable to that of the high-frequency intralayer vibrational modes, especially the peak at 12.5

cm$^{-1}$ (0.35 THz), indicating strong coupling between the excitons and the LB modes in pentagonal PdSe$_2$.

Coincident with the Raman results at 144.6 cm$^{-1}$, we assign the periodic signal at 4.3 THz as the signature of coherent phonons with intralayer atomic vibrations within an individual PdSe$_2$ layer. A possible explanation is that the intralayer atomic deformations can dynamically alter the interatomic electron cloud over femtosecond time scales, and therefore give rise to a transiently modified permittivity, making it detectable as a frequency modulated differential transmission contrast with our high time-resolvability spectroscopy. Benefited with the broadband probe, the phonon coupling with different excitonic transitions, at both temporal and spectral domains, can be compared simultaneously. Without any coupling, the oscillatory amplitude and coherence lifetime would be zero. Under a broadband probe with a fixed excitation strength, different types of the collective oscillations of atoms are initiated and then become decoherent at the same level, but they have different responses at different transitions. For transitions with stronger coupling with phonons, more pronounced oscillatory amplitudes and longer lasting coherence lifetimes are expected. As shown in Fig. 3b, the 4.3 THz intensity modulation $\Delta T/T$ related to excited-state absorption and carrier absorption (1.2-1.8 eV) is more pronounced than that near the vicinity of the exciton resonance. With the intralayer coherent atomic motions coupling preferentially to carriers, they influence the strength of carrier-induced screening and considerably alter the electronic bandgap at an extremely high frequency of 4.3 THz, as discussed earlier in Fig. 3d. The observed coherent phonon oscillations at 0.35 THz can be assigned to the low-frequency LB$_{8,7}$ mode originating from the out-of-plane interlayer vibrations and such strong coupling is in good agreement with the large Raman intensity of LB$_{8,7}$ peak. This can be attributed to the modified permittivity based on the coherent modification of the electronic structure. As shown in Fig. 4b, the 0.35 THz oscillation couples most strongly near the exciton resonance, while the coupling is much weaker in the regime of excited-state absorption. In contrast to the absence of coherent signals in the PB band of black phosphorus[17], the strong coherent phonon oscillation of PdSe$_2$ at 0.35 THz preferentially couples with excitons at the PB band, indicating the exceptional strong exciton−phonon coupling in PdSe$_2$.

**Electron−phonon coupling from first-principles**

To confirm our picture of phonon dynamics in PdSe$_2$, we perform phonon and electron−phonon coupling calculations using the finite difference method[42]. Figure 6a demonstrates the calculated phonon dispersion of eight-layer PdSe$_2$, showing this configuration to be dynamically stable. The phonon occupation number $N_{ph}$ is obtained from the Bose–Einstein distribution function at 300 K. The calculated displacement patterns in Fig. 6b confirm that the higher frequencies around 4.3 THz (144.6 cm$^{-1}$) are dominated by intralayer vibrations of Se atoms, whilst the low-frequency modes below 1 THz (33.3 cm$^{-1}$) are dominated by interlayer vibrations of both Pd and Se atoms. In the low-frequency regime, the calculated phonon frequencies are lower than the measured ones, because the generalized gradient approximation to the exchange correlation functional tends to underestimate the interatomic bonding. Therefore, the calculated LB$_{8,7}$ frequency of 0.30 THz is accordingly lower than the measured frequency of 0.35 THz. The strength of electron-phonon coupling is estimated using a frozen-phonon approach along the normal modes with frequencies around 4.3 and 0.35 THz at the Γ point. It is found that there are two modes with much stronger coupling strength than their neighboring modes (Supplementary Fig. 5). As shown in Fig. 6d, the high-frequency intralayer mode at 4.30 THz significantly modulates the high-lying conduction bands around 3 eV. This is because the high-lying conduction bands come from the intralayer orbitals along the Pd–Se bonds (Fig. 6c), and only strong intralayer displacements can influence their wave function. By contrast, the interlayer mode of 0.30 THz strongly couples with the highest valence bands. These bands, consisting mainly of $d_{z^2}$ orbitals of Pd and $p_z$ orbitals of Se, are the dominant hole states in the formation of the A exciton. The $d_{z^2}$ orbitals of Pd and $p_z$ orbitals of Se, oriented along the out-of-plane direction, are extremely sensitive to the interlayer breathing mode, which strongly couples to their charge distribution (Fig. 6c). The calculated electron−phonon dynamics are consistent with the observed ultrafast periodic modulation of both $\Delta T/T$ and $\Delta E$ in our broadband pump-probe experiments, in which the high energy carriers are more sensitive to the intralayer mode, while the A exciton is more sensitive to the interlayer mode. It should be noted that an accurate estimation of the different coupling processes requires detailed computation of both electron-phonon and exciton-phonon coupling strengths[43], but here we use the frozen-phonon method of selected modes motivated by the experimental results (for the variation of exciton absorption under these two phonon modes, see Supplementary Fig. 7).

In conclusion, our findings provide a pathway for the future development of THz frequency optoelectronic devices that based on layered PdSe$_2$. Furthermore, the observed vibrational phenomena combined with theoretical calculations provide an intuitive picture for exciton–phonon and electron–phonon interactions in 2D layered materials, which is supported by the ultralow frequency Raman spectroscopy with frequencies down to 5 cm$^{-1}$.

**METHODS**

**Sample preparation**
High-quality PdSe$_2$ samples are grown on quartz substrate directly by chemical vapor deposition (CVD) method using PdCl$_2$ and Se powder as precursors. Selenium is evaporated at 250°C, PdCl$_2$ is in the middle of a three-zone tube furnace with an intermediate temperature of 500°C, and quartz substrate is in the last high-temperature zone of 600°C. The evaporated Se and $PdCl_2$ are grown on a quartz substrate in a high-temperature zone dragged by 300 standard cubic centimeters per minute (sccm) Ar and 30 sccm $H_2$. The furnace is kept at 600°C for 20 minutes. Then, the temperature is cooled down to room temperature.

**Femtosecond transient absorption spectroscopy**

The broadband transient absorption spectroscopy utilized a Ti:sapphire ultrafast amplifier (Legend Elite-1K-HE Coherent, 800nm, ~35 fs, ~4mJ, and 1 kHz) and pump-probe transient absorption spectrometer (Helios, Ultrafast System). The generated laser pulses are split into two optical paths by a pellicle beam splitter. One weak portion of the laser beam is used to generate white light continuum as the broadband probe after passing through a sapphire or $CaF_2$ crystal. Both the supercontinuum crystal and photodetectors need to be switched for the visible and near-infrared band. The other strong pulse is used to pump a nonlinear optical parametric amplifier (Light Conversion TOPAS) to produce pump laser pulses with tunable photon energy. The resolution of the pump–probe time delay is ~20 fs, tuned by a delay line (Thorlabs). The pump–probe time delay is controlled by a highly accurate motorized optical delay line. The pump pulse is modulated at 500 Hz by a mechanical chopper to ensure that the signal with or without photoexcitation can be recorded alternately. In this work, the pump and probe polarizations are controlled by a 1/2 plate and are crossed at the sample. The polarizations of the two pulses are aligned in such a way that the angle between them is a magic angle of 54.7° to avoid any potential artifacts due to rotation effects. All acquired data are corrected by the frequency-dependent chirp and background scattered light.

**Electronic structures calculations**

First-principles calculations are performed on the basis of density functional theory (DFT) with the projector augmented wave (PAW) method as implemented in the Vienna *ab initio* simulation package (VASP)[44]. The Perdew–Burke–Ernzerhof (PBE) parameterization of the generalized gradient approximation (GGA) is adopted as the exchange-correlation functional[45]. To describe van der Waals interactions in eight-layer $PdSe_2$, the Tkatchenko-Scheffler method is used, which provides better agreement with the measured lattice constants compared to other methods[46]. A kinetic cutoff energy of 500 eV with an energy threshold of $10^{-6}$ eV for the self-consistent cycle is employed. The Brillouin zone is sampled using the Monkhorst-Pack method with a **k**-point grid of size 5×5. The cell geometry and atomic positions are fully optimized using the conjugate gradient (CG) algorithm until the maximum force on each unconstrained atom is less than 0.01 eV/Å. We maintain an interlayer vacuum spacing larger than 16 Å to eliminate interactions between adjacent slabs. Spin−orbit coupling is found to have negligible influence on the electronic structure. Many body perturbation theory calculations are conducted to obtain the quasiparticle electronic structures, in which the electron-electron interactions are included in the self-energy[47,48]. The single shot $G_0W_0$ method is applied on top of the HSE06 hybrid functional[38]. Careful convergence tests are performed using **k**-point grids of 6×6, 10×10 and 12×12. Consistent with previous work, a 6×6 **k**-point grid can produce a qualitatively good dielectric function comparable to that of the 14×14 grid[49], and a 10×10 **k**-point grid is well-converged. In addition, the $G_0W_0$ dielectric functions calculated using 768 and 384 bands show no significant difference. Therefore, we use 384 bands with a 10×10 **k**-point grid in the calculations. Many body corrections are included in the band structure in Figure 1f in the form of a scissor shift of 0.916 eV.

**Lattice dynamics and electron-phonon coupling calculations**

We obtain the harmonic interatomic force constants using the finite displacement method in a 2×2 supercell with the PHONOPY code[50]. The full phonon dispersion is shown in Supplementary Fig. 6. We estimate the strength of electron−phonon coupling by calculating the electronic structure along the normal mode $u_{\mathbf{q}\nu}$[41]:

$$u_{\mathbf{q}\nu} = \frac{1}{\sqrt{N_p}} \sum_{\mathbf{R}_p,\alpha,i} \sqrt{m_\alpha} h_{p\alpha i} e^{-i\mathbf{q}\cdot\mathbf{R}_p} w_{-\mathbf{q}\nu;i\alpha} \quad (1)$$

where $\mathbf{q}$ and $\nu$ are reciprocal space phonon wavevector and branch respectively, $N_p$ is the number of primitive cells in the real space supercell, $\mathbf{R}_p$ is the position vector of unit cell $p$, $m_\alpha$ is the nuclear mass of atom $\alpha$, $i$ runs over Cartesian coordinates, $h_{p\alpha i}$ is the displacement coordinate, and $w_{\mathbf{q}\nu;i\alpha}$ is the corresponding eigenvector. We use an amplitude of $u_{\mathbf{q}\nu}$ corresponding to the 1$\sigma$ of the Gaussian describing the ground state phonon wave function, given by $1/\sqrt{2\omega_{q\nu}}$, where $\omega_{\mathbf{q}\nu}$ is the phonon frequency. We average over negative and positive displacements to calculate the absolute energy change in the presence of the frozen phonons. The electron-phonon coupling strengths for different phonon modes at the Γ point are shown in Supplementary Fig. 5, which can be directly compared with time- and angular-resolved photoelectron spectroscopy (TR-ARPES)[51,52].

## DATA AVAILABILITY

The authors declare that all data supporting the findings of this study are available within the paper and its supplementary information files.

## AUTHOR INFORMATION

**Corresponding Author**

*E-mails: drfchen@sdu.edu.cn; bm418@cam.ac.uk; phtan@semi.ac.cn.

**COMPETING INTERESTS**

The authors declare that there are no competing interests.

**ACKNOWLEDGMENT**

This work is supported by the National Natural Science Foundation of China (NSFC) (11535008, 11874350); The 111 Project of China (No. B13029); CAS Key Research Program of Frontier Sciences (Grant No. ZDBS-LY-SLH004). B.P. and B.M. acknowledge support from the Winton Programme for the Physics of Sustainability, and B.M. also acknowledges support from the Gianna Angelopoulos Programme for Science, Technology, and Innovation. The calculations were performed using resources provided by the Cambridge Tier-2 system operated by the University of Cambridge Research Computing Service (http://www.hpc.cam.ac.uk) and funded by EPSRC Tier-2 capital grant EP/P020259/1, and also with computational support from the UK Materials and Molecular Modelling Hub, which is partially funded by EPSRC (EP/P020194), for which access is obtained via the UKCP consortium and funded by EPSRC grant ref EP/P022561/1.

**AUTHOR CONTRIBUTIONS**

L.Z. and F.C. conceived and designed the project. L.Z. performed the femtosecond pump-probe experiments, analysed the data and wrote the manuscript. B.P. and B.M conducted the theoretical part and wrote the corresponding part of the manuscript. Y.C.L., M.L.L., and P.H.T. performed and interpreted the Raman data. L.Z., B.Z., and C.P. performed the sample characterizations. All authors have participated in the interpretation of the experiments and contributed to the manuscript.

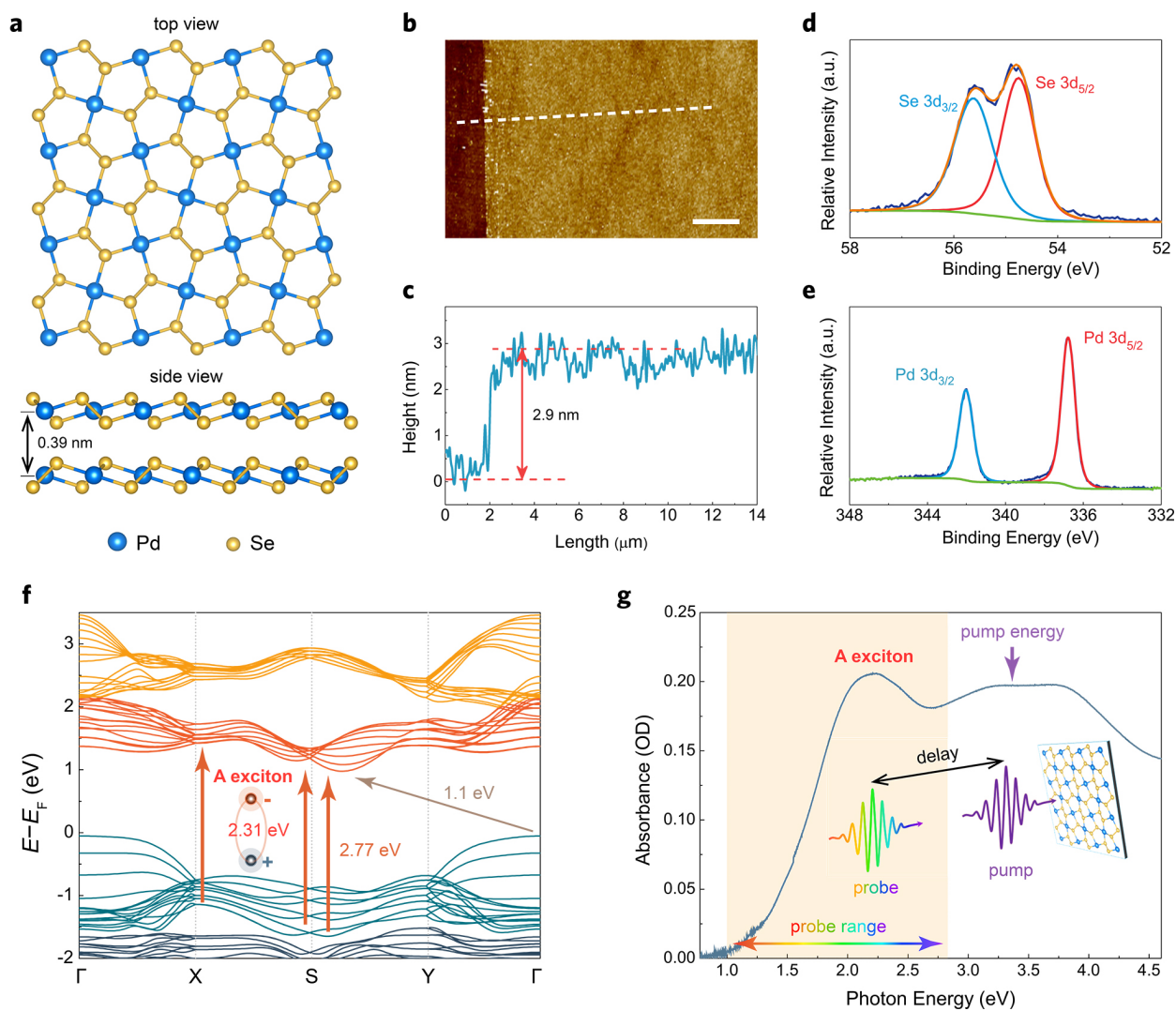

**Fig. 1. Steady-state characterizations of PdSe$_2$. a** Top and side views of PdSe$_2$ in Cairo pentagonal tiling pattern with puckered structure. The Se and Pd atoms are shown in pale gold and blue, respectively. **b** Nanoscale surface topographic image of the as-synthesized PdSe$_2$ on quartz substrate. The white scale bar is 3 μm. **c** Height profiles derived from the AFM image. High-resolution XPS spectrum of **d** Se 3*d* and **e** Pd 3*d*. **f** Electronic band structure of eight-layer PdSe$_2$ obtained from first-principles calculations. **g** Measured equilibrium absorption spectrum of the sample. The inset is the schematic illustration of pump-probe experiments.

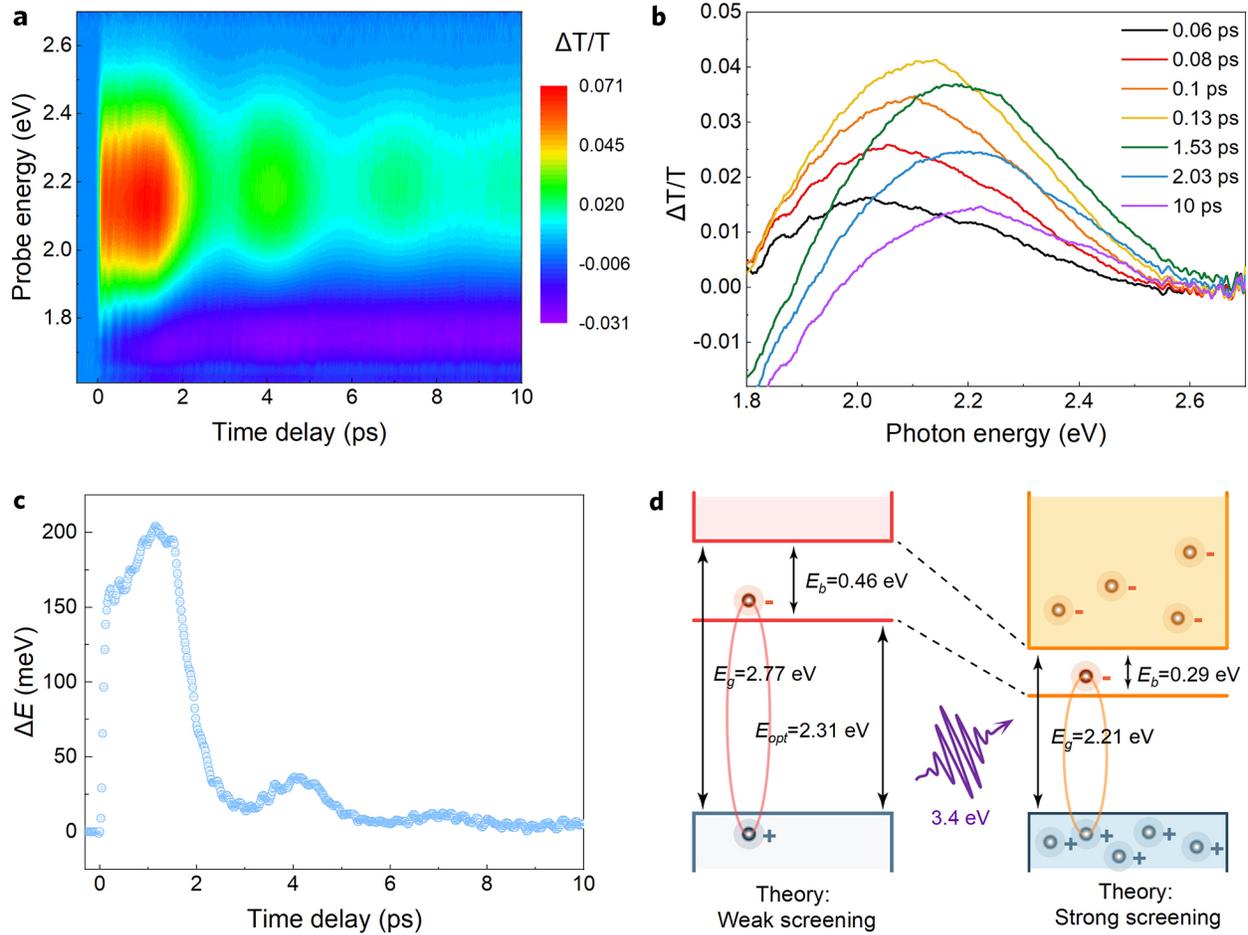

**Fig. 2. Femtosecond-laser-driven spectral and temporal evolution of PdSe$_2$. a** 2D transient absorption map as a function of probe photon energy and pump-probe time delay. **b** Transient $\Delta T/T$ spectra of PdSe$_2$ sample at different pump-probe time delays. **c** Extracted exciton energy shift ($\Delta E$) as a function of pump-probe time delay. **d** Schematic of the electronic band structure at weak and strong screening regimes. The overall $\Delta E$ results from an interplay between the quasiparticle bandgap ($E_g$) shrinkage and the exciton binding energy ($E_b$) reduction due to the presence of photoexcited carriers.

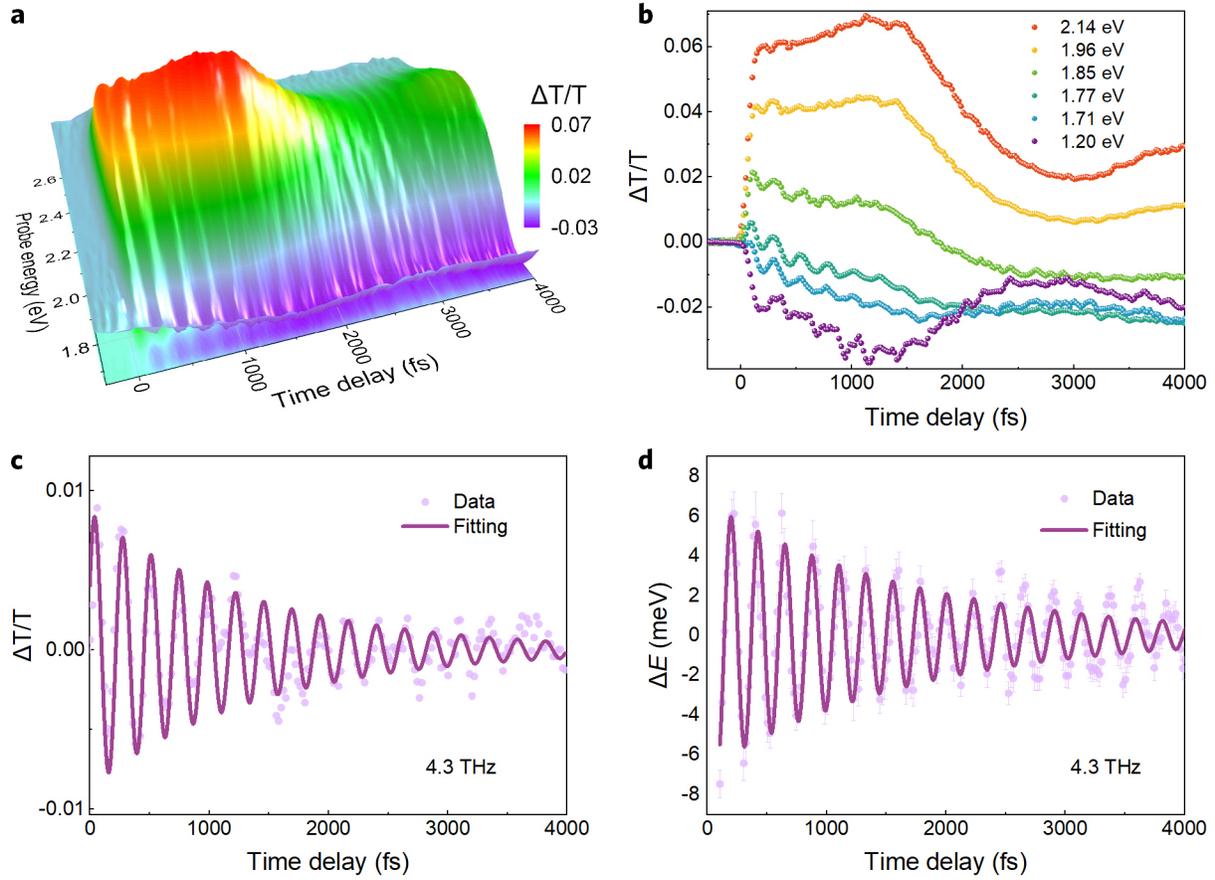

**Fig. 3. Real-time observation of intralayer coherent vibrational dynamics of PdSe$_2$. a** 3D transient absorption spectrum in terms of differential transmission $\Delta T/T$ as a function of time delay and probe photon energy on femtosecond timescale. **b** The corresponding time-dependent dynamic process with different probe photon energies. **c** Coherent phonon dynamics at 4.3 THz probed at 1.71 eV. The oscillatory components of transient absorption spectrum are extracted after subtracting the population dynamics. **d** Extracted oscillatory components of $\Delta E$ as a function of pump-probe time delay. The fitted lifetime is 1.73 ps.

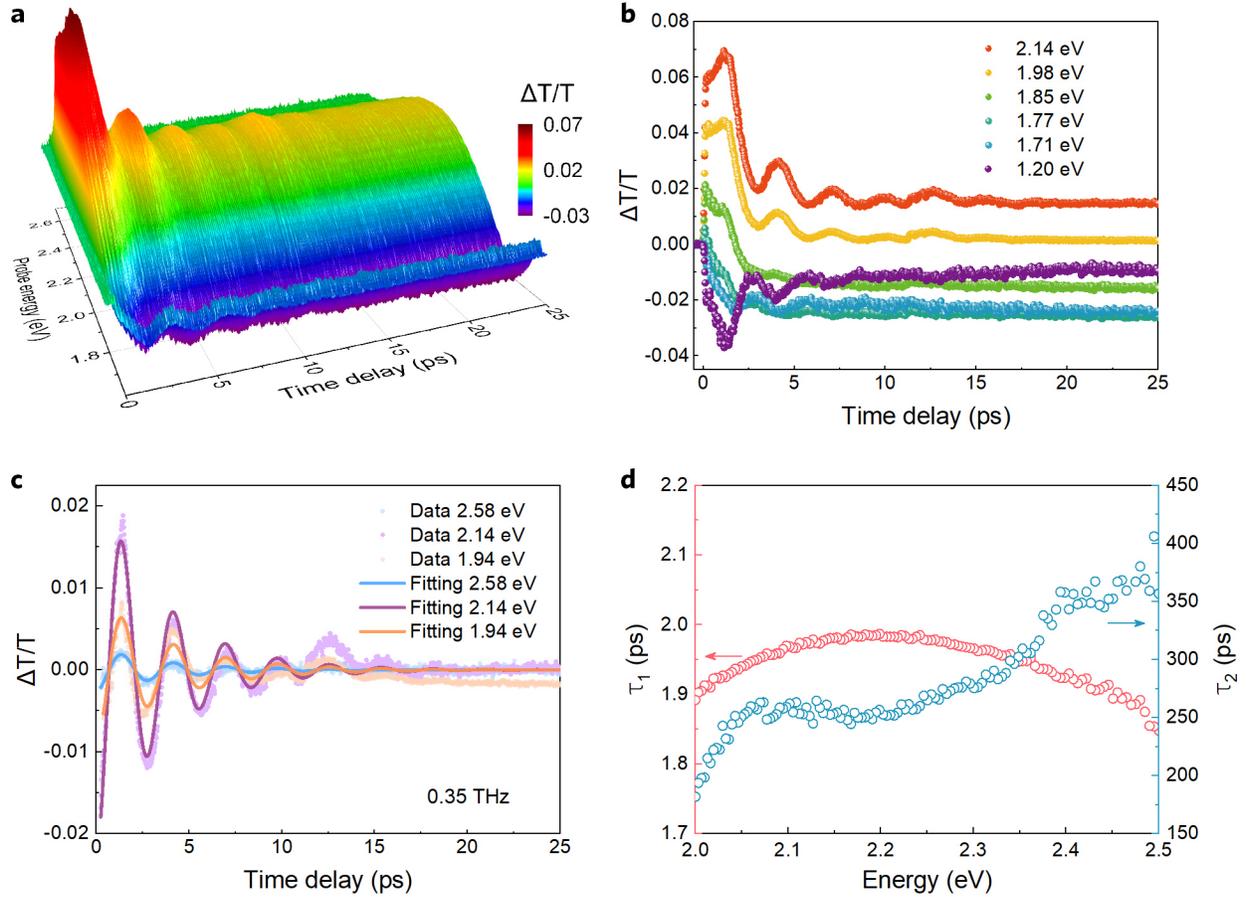

**Fig. 4. Real-time observation of interlayer coherent vibrational dynamics of PdSe$_2$. a** 3D transient absorption spectrum in terms of differential transmission $\Delta T/T$ as a function of time delay and probe photon energy on picosecond timescale. **b** Time-dependent $\Delta T/T$ with different probe photon energies. **c** Coherent phonon dynamics at 0.35 THz probed at photon energies of 2.58, 2.14, and 1.94 eV. The vibrational traces are fitted using an exponentially damped sinusoid model. **d** Fast decay time $\tau_1$ and exciton lifetime $\tau_2$ as a function of probe photon energy. The best-fit value is fitted using bi-exponential decay model with a criterion of minimum standard deviation of the residuals.

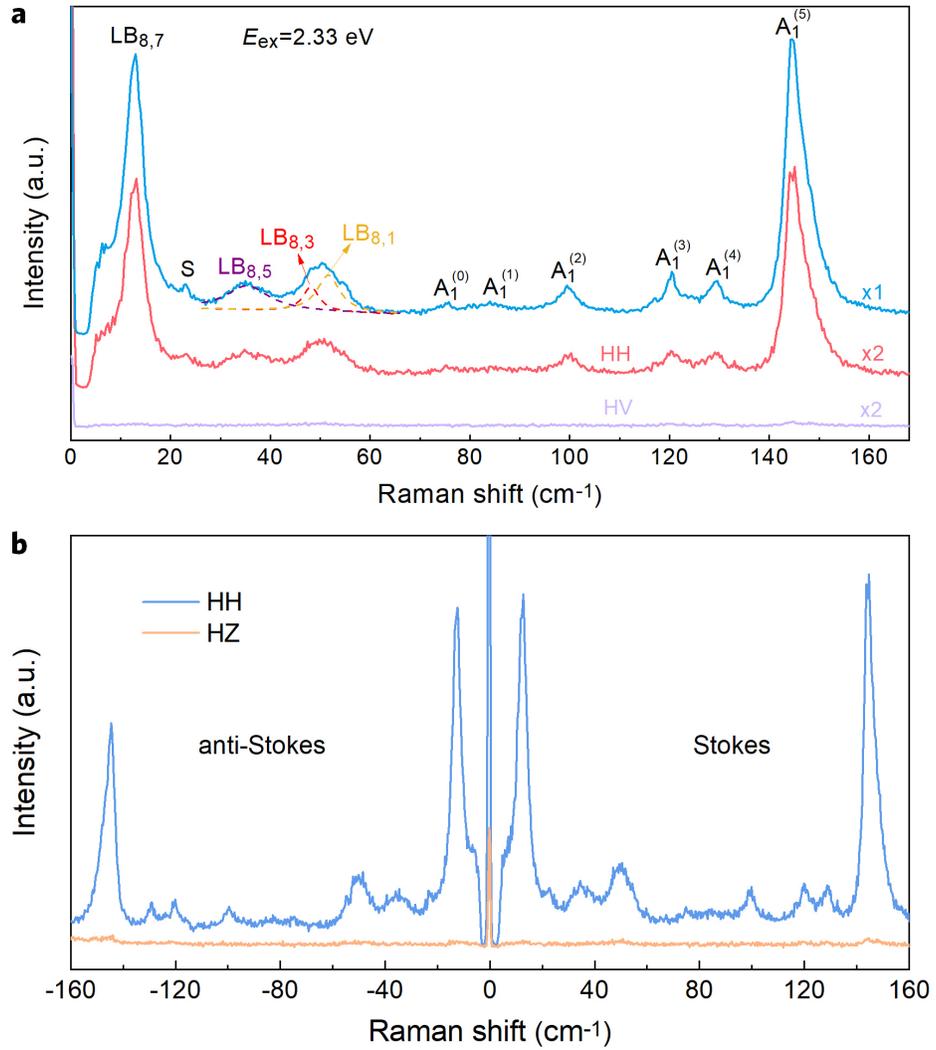

**Fig. 5. Ultralow-frequency Raman spectroscopy of PdSe$_2$. a** Raman spectra of under unpolarized, parallel (HH) and cross (HV) polarization configurations. **b** Stokes/anti-Stokes Raman spectra measured in HH and HV polarization configurations.

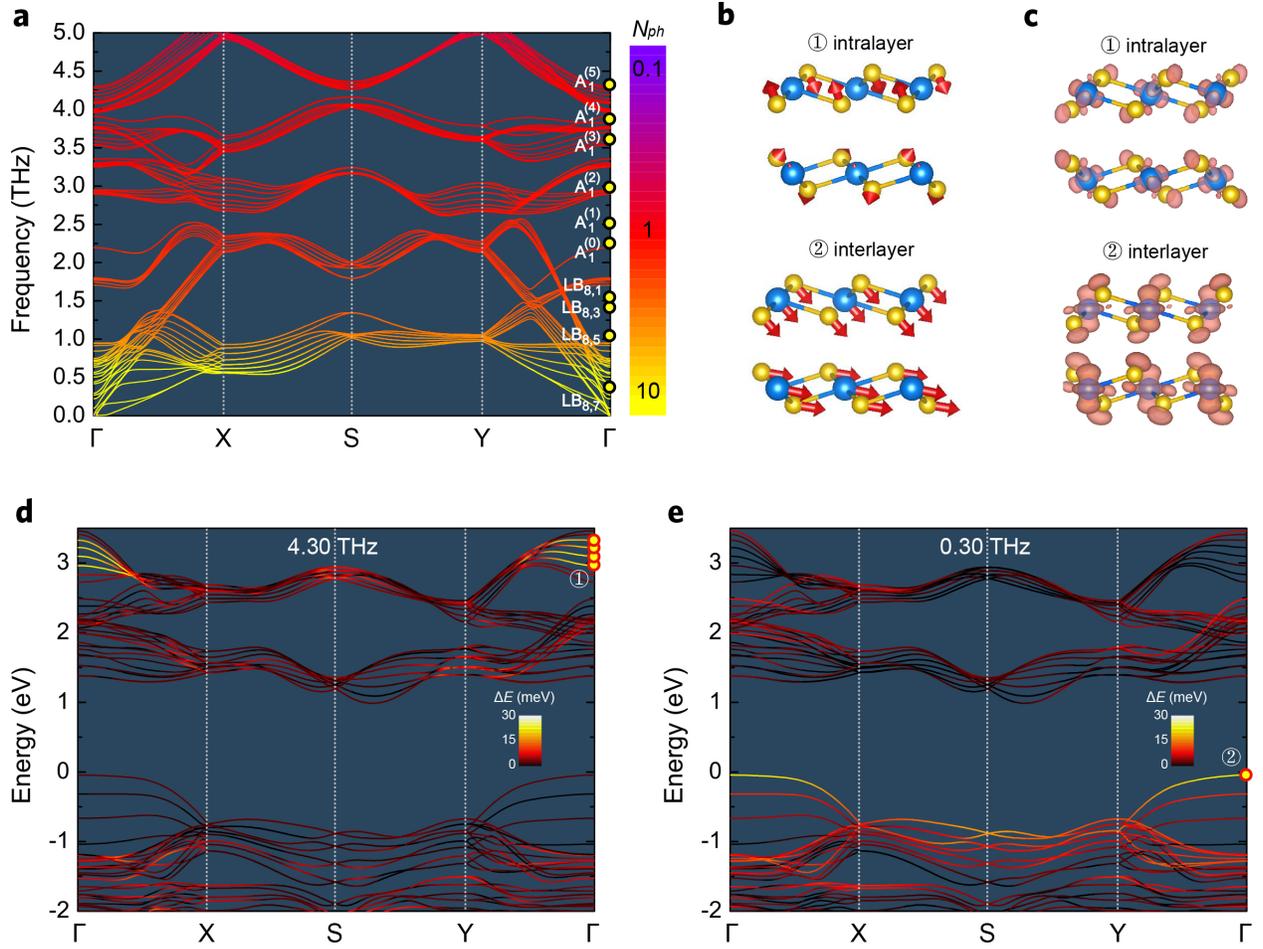

**Fig. 6. Lattice dynamics and electron-phonon coupling in PdSe$_2$. a** Calculated phonon dispersion spectrum of eight-layer PdSe$_2$. **b** Displacement patterns of the intralayer and interlayer modes. **c** Partial charge density of the electronic states that couple most strongly to the corresponding atomic displacements. Electronic structure modulated by **d** intralayer mode at 4.30 THz and **e** interlayer mode at 0.30 THz.